\documentclass[fleqn,10pt]{article}
\usepackage{epsfig}
\newcommand{\be}{\begin{equation}}
\newcommand{\ee}{\end{equation}}

\newcommand{\eq}[1]{Eq.~(\ref{#1})}
\newcommand{\fig}[1]{Fig.~\ref{#1}}
\newcommand{\chii}{\chi_{{}_{I}}}
\newcommand{\chir}{\chi_{{}_{R}}}

\def\lpa{\lambda_{p{-}\rm air}}
\def\spa{\sigma_{p{-}\rm air}}
\def\spai{\sigma_{p{-}\rm air}^{\rm inel}}
\def\spae{\sigma_{p{-}\rm air}^{\rm el}}
\def\spaqe{\sigma_{p{-}\rm air}^{q{-}\rm el}}

\begin{document}    
\setcounter{secnumdepth}{4}
\renewcommand\thepage{\ }
%
%
\begin{titlepage} 
%
\newcommand\reportnumber{708} 
\newcommand\mydate{March, 2000} 
\newlength{\nulogo} 
\settowidth{\nulogo}{\small\sf{N.U.H.E.P. Report No. \reportnumber}}
\title{
\vspace{-.8in} 
\hfill\fbox{{\parbox{\nulogo}{\small\sf{Northwestern University: \\
N.U.H.E.P. Report No. \reportnumber\\
          \mydate}}}}
\ \vspace{0.5in} \\
{
Breaking the Barriers---Uniting Accelerator and Cosmic Ray p-p Cross Sections
}}
 
\author{
M.~M.~Block
\thanks{Work partially supported by Department of Energy contract
DA-AC02-76-Er02289 Task B.}\vspace{-5pt}   \\
{\small\em Department of Physics and Astronomy,} \vspace{-5pt} \\ 
{\small\em Northwestern University, Evanston, IL 60208}\\
\vspace{-5pt}\\
%
F.~Halzen
\thanks{Work partially supported by
Department of Energy contract
DE-AC02-76ER0088 and
the University of Wisconsin Research
Committee with funds granted by the Wisconsin Alumni Research Foundation.}
\vspace{-5pt} \\
{\small\em Department of Physics,} \vspace{-5pt} \\
{\small\em University of Wisconsin, Madison, WI 53706}  \\
\vspace{-5pt}\\
%
G.~Pancheri
\vspace{-5pt} \\
{\small\em INFN-Laboratori Nazionali di Frascati,} \vspace{-5pt} \\ 
{\small\em Frascati, Italy}\\
\vspace{-5pt}\\
T. Stanev\thanks{Work partially supported by the U.S.~Department of Energy under Grant No. DE-FG02-91ER40626.}
\vspace{-5pt} \\
{\small\em Bartol Research Institute, University of Delaware, Newark, DE 19716
}
\vspace{.5in}\\
{\small \sf Paper presented by Martin M. Block}\\[-5pt]
{\small  {\it mblock@nwu.edu}}\\[-6pt]
{\small \sf  at the}\\[-6pt]
{\small \sf 25th Pamir-Chacaltaya Collaboration Workshop, Lodz, Poland, Nov. 3-7, 1999}\\[-6pt]
\vspace{-5pt}\\
%
}    
%
\vfill
\date{} 
\maketitle
\begin{abstract}
We make a QCD-inspired parameterization
 of all accelerator data on forward proton-proton and antiproton-proton
 scattering amplitudes.  Using vector dominance and the additive quark model, we show that the same parameters also fit $\gamma p$ and $\gamma \gamma$ interactions. Using the high energy predictions of our model, along with Glauber theory,  we calculate proton--air
cross sections at energies near $\sqrt s \approx$ 30 TeV.  The comparison of p-air cosmic ray measurements with our QCD model predictions provide a strong constraint on the inclusive particle production cross section.

\end{abstract}
\end{titlepage} 
\pagenumbering{arabic}
\renewcommand{\thepage}{-- \arabic{page}\ --}  
%
\section{Introduction}
This communication is divided into three sections.  

First, we show that the data on the total cross section, the
slope parameter $B$ of the elastic differential cross section, and the
ratio of the real to imaginary part of the forward scattering
amplitude $\rho$ for $pp$ and $\bar p p$ interactions can be nicely
described by a model where high energy cross sections grow with energy
as a consequence of the increasing number of soft partons populating
the colliding particles \cite{margolis},\cite{blockprd}. The differential cross
sections for the Tevatron and LHC are predicted. 

Next, we verify the model by showing that the known experimental data on $\gamma p$ and $\gamma\gamma$ interactions 
can be derived from our $pp$ and $\bar p p$ forward scattering
amplitudes using vector meson dominance (VMD) and the additive quark model\cite{blockprd}.

Finally, we use the high energy predictions of our QCD-inspired parameterization
of accelerator data on forward proton-proton and antiproton-proton
scattering amplitudes, along with Glauber theory, to predict proton--air
cross sections at energies near $\sqrt s \approx$ 30 TeV\cite{pair}. 

All cross sections will be computed in an eikonal formalism
guaranteeing
unitarity throughout:
\begin{eqnarray}
\sigma_{tot}(s)&=&2\int\,\left\{1-e^{-\chii (b,s)}\cos[\chir(b,s)]
\right\}\,d^2\vec{b}.
\end{eqnarray}
Here, $\chi $ is the complex eikonal ($\chi=\chir + i\chii$), and
$b$ is the impact parameter.  The even eikonal profile function
$\chi^{even}$ receives contributions from quark-quark, quark-gluon
and gluon-gluon interactions, and therefore
\begin{eqnarray}
\chi^{even}(s,b) &=& \chi_{qq}(s,b)+\chi_{qg}(s,b)+\chi_{gg}(s,b)
\nonumber \\
&=& i\left[ \sigma_{qq}(s)W(b;\mu_{qq})
+ \sigma_{qg}(s)W(b;\sqrt{\mu_{qq}\mu_{gg}})
+ \sigma_{gg}(s)W(b;\mu_{gg})\right]\, ,
\label{chiintro}
\end{eqnarray}
where $\sigma_{ij}$ are the cross sections of the colliding partons,
and $W(b;\mu)$ their overlap function in impact parameter space,
parameterized as the Fourier transform of a dipole form factor.  
The impact parameter space distribution function
\begin{equation}
W(b\,;\mu)=\frac{\mu^2}{96\pi}(\mu b)^3K_3(\mu b)\label{W}
\end{equation}
is normalized so that $\int W(b\,;\mu)d^2 \vec{b}=1$. As a consequence
of both factorization and the normalization chosen for the
$W(b\,;\mu)$, it should be noted that
\be
\int \chi^{even}(s,b)\, d^2\vec b=i\left[\sigma_{gg}(s)
+\sigma_{qg}(s)+\sigma_{qq}(s)\right],
\label{integrateeven}
\ee
so that $\sigma_{tot}^{even}(s)=2
\,{\rm Im}\left\{ i\left[\sigma_{
gg}(s)+\sigma_{qg}(s)+\sigma_{qq}(s)\right]\right\}$,
for small $\chi$.
This
formalism is identical to the one used in ``mini-jet" models
\cite{minijet}, as well as in simulation programs for minimum-bias
hadronic interactions such as PYTHIA and SIBYLL\cite{minijet}.

In this model hadrons asymptotically evolve into black disks of
partons. The rising cross section,
asymptotically associated with gluon-gluon interactions, is simply
parameterized by a normalization, an energy scale, and two parameters:
$\mu_{gg}$ which describes the ``area" occupied by gluons in the
colliding hadrons, and $J (= 1+\epsilon)$. Here, $J$ is defined via
the gluonic structure function of the proton, which is assumed to
behave as $1/x^J$ for small x. It therefore controls the soft gluon
content of the proton. The introduction of the quark-quark and
quark-gluon terms allows us to adequately parameterize the data at all
energies, since the ``size'' of quarks and gluons in the proton can be
different. In the present context, this model represents a convenient
parameterization of the $pp$ and $p\bar{p}$ forward scattering
amplitude.

The photoproduction cross sections are calculated from this
parameterization assuming vector meson dominance and the additive
quark model. For the probability that the photon interacts as a hadron
($P_{had}$), we use the value $P_{had}=1/240$ which can be derived
from vector meson dominance. Our results show that its value is indeed
independent of energy. It is, however, uncertain by 20\% because it
depends on whether we relate photoproduction to $\pi$-nucleon or
nucleon-nucleon data (In other words, $\pi N$ and $NN$ total cross
sections only satisfy the additive quark model to this accuracy).
Subsequently, following reference
\cite{fletcher}, we obtain $\gamma p$ cross sections from the
assumption that, in the spirit of VMD, the photon is a 2 quark state
in contrast with the proton which is a 3 quark state. The $\gamma p$
total cross section is obtained from the even eikonal for $pp$ and
$\bar p p$ by the substitutions $\sigma_{ij} \rightarrow
\textstyle\frac{2}{3} \,\sigma_{ij}$ and $\mu_i \rightarrow
\sqrt{\textstyle\frac{3}{2}} \,\mu_i$.

We will thus produce a parameter-free description of the total
photoproduction cross section, the phase of the forward scattering
amplitude and the forward slope for $\gamma p \rightarrow  Vp$,
where $V= \rho, \omega, \phi$. Interestingly, our results on the phase
of $Vp \rightarrow Vp$ are in complete agreement with the values
derived from Compton scattering results ($\gamma + p \rightarrow
\gamma + p$) using dispersion relations. We also calculate the total
elastic and differential cross sections for $\gamma p \rightarrow
Vp$. This wealth of data is accommodated without discrepancy.

The $\gamma\gamma$ cross sections are derived following the same
procedure. We now substitute $\sigma_{ij} \rightarrow
\textstyle\frac{4}{9} \,\sigma_{ij}$ and $\mu_i \rightarrow
\textstyle\frac{3}{2}\,\mu_i$ into the nucleon-nucleon even eikonal,
and predict the total cross section and the differential cross
sections for all reactions $\gamma\gamma \rightarrow V_i V_j$ at a
variety of energies, where $V= \rho, \omega, \phi$.

The high energy $\gamma\gamma$ total cross section \cite{exp} have
been measured by two experiments at LEP. While these measurements
yield new information on its high energy behavior at center-of-mass
energies in excess of $\sqrt{s}=15$ GeV, they may represent the last
opportunity to measure the $\gamma\gamma$ cross section, and the two
data sets appear to disagree. However, it has been argued that the
original data are consistent within the errors \cite{vancouver} and
that the observed disagreements are due to two different Monte Carlo's
used to extract the quoted values. We here point out that our analysis
nicely accommodates the L3 result \cite{previous}. Our model
approximately satisfies the factorization theorem,
$\sigma_{pp}/\sigma_{\gamma p} = \sigma_{\gamma p}/\sigma_{\gamma
\gamma}$, because of its small eikonal.  The OPAL data do not satisfy
it. In fact, no model incorporating the additive quark model and
factorization can accommodate the OPAL data. VMD and factorization are
sufficient to prevent one from adjusting $P_{had}$, or any other
parameters, to change this conclusion.
\section{High energy proton-proton and proton-antiproton scattering}
In this section we discuss our QCD-inspired parameterization of the
forward amplitudes. To determine its parameters, we fit
all high energy forward $\bar pp$ and $pp$ scattering data
above 15 GeV, for the total cross section ($\sigma_{tot}$), the ratio
of the real to the imaginary part of the forward scattering amplitude
($\rho$), and the logarithmic slope of the differential elastic
scattering cross section in the forward direction ($B$). Then, we
compare the experimental data for the elastic scattering cross
section and for the differential elastic scattering with our
results. Finally, a prediction is made for the differential elastic
scattering at the LHC.

Our QCD-inspired parameterization satisfies crossing symmetry, {\em
i.e.,} it is either even or odd under the transformation $E\rightarrow
-E$, where $E$ is the laboratory energy. This allows us to
simultaneously describe $\bar pp$ and $pp$ scattering. It also
satisfies analyticity, and unitarity because of the eikonal formalism.
Since the total cross section asymptotically rises as $\log^2s$, our
QCD-inspired parameterization complies with the Froissart bound.  The
eikonal formalism for calculating $\sigma_{tot}$, $\rho$ and $B$, along
with details on the analyticity, the
Froissart bound, and the QCD-inspired eikonal are given in ref. \cite{blockprd}.  
In all 11 parameters are used. The low energy region, where
the differences between $\bar p p$ and $pp$ scattering are
substantial, largely determines the 7 parameters necessary to fit the
odd eikonal and the quark-quark and quark-gluon contribution to the
even eikonal. Thus, they largely decouple from the high energy
behavior.  Hence, for $\sqrt s\ge 25$ GeV, where the difference
between $\bar p p$ and $pp$ scattering becomes small, only 4
parameters describe all data.

\begin{figure}[htbp]
\begin{center}
\mbox{\epsfig{file=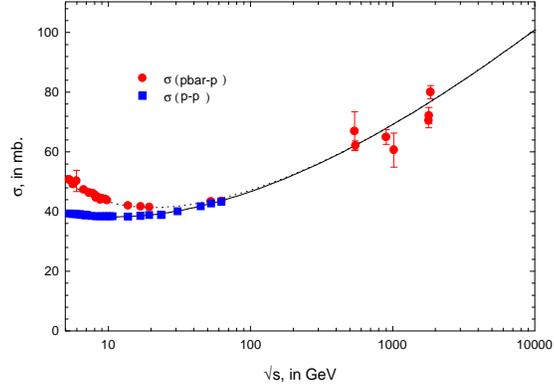,width=3.5in,%
bbllx=50pt,bblly=335pt,bburx=565pt,bbury=685pt,clip=}}
\end{center}
\caption[]{The total cross section for $pp$ and $\bar p p$
scattering.  The solid line and squares are for $pp$ and the dotted
line and circles are for $\bar p p$.}
\label{fig:1sigtot}
\end{figure}

\begin{figure}[htp]
\begin{center}
\mbox{\epsfig{file=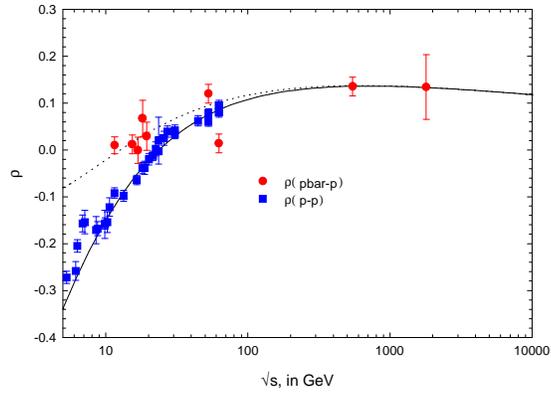,width=3.5in,%
bbllx=50pt,bblly=335pt,bburx=565pt,bbury=685pt,clip=}}
\end{center}
\caption[]{The ratio of the real to imaginary part of the
forward scattering amplitude for $pp$ and $\bar p p$ scattering. The
solid line and squares are for $pp$ and the dotted line and circles
are for $\bar p p$.}
\label{fig:1rho}
\end{figure}

\begin{figure}[htbp] 
\begin{center}
\mbox{\epsfig{file=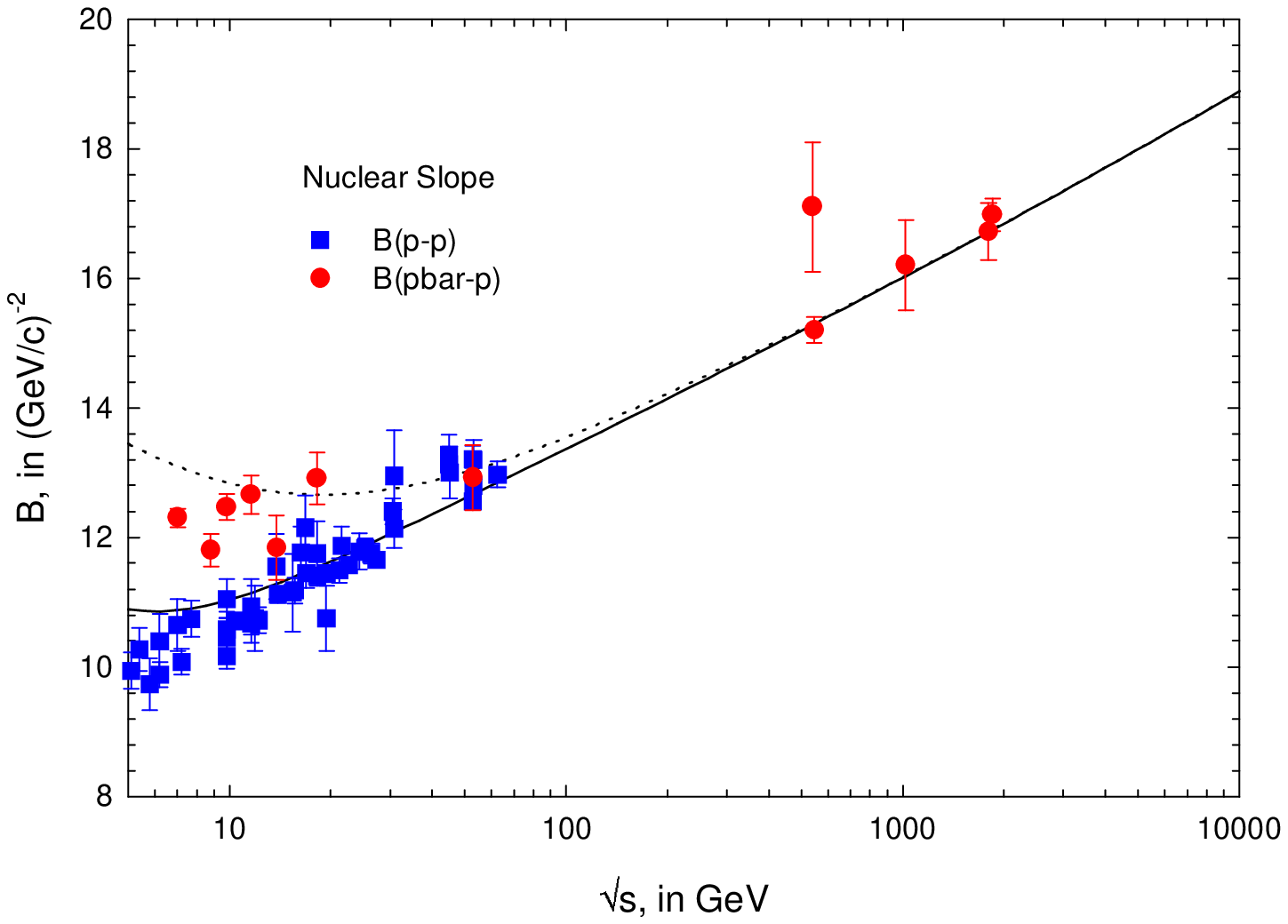,width=3.5in,%
bbllx=50pt,bblly=335pt,bburx=565pt,bbury=685pt,clip=}}
\end{center}
\caption[]{The nuclear slope parameter
for elastic $pp$ and $p\bar p$ scattering.  The solid line and squares
are for $pp$ and the dotted line and circles are for $\bar p p$.}
\label{fig:1b}
\end{figure}

We fit all the highest energy cross section data (E710
\cite{E710}, CDF \cite{CDF} and the unpublished Tevatron value
\cite{newpp}), which anchor the upper end of our cross section curves.
The results of the fit are shown in Fig.~\ref{fig:1sigtot}. Data for
$\rho$ values and $B$ are confronted with our model in Figs.\
\ref{fig:1rho} and \ref{fig:1b}.

It can be seen from those figures that we obtain a satisfactory
description of all 3 quantities, for both $\bar pp$ and $pp$
scattering. The $\chi^2$ of the fit is reasonably good (considering
the large spread in some of the experimental data, as well as the
discrepancies in the highest energy cross sections), giving a
$\chi^2/d.f.=1.66$, for 75 degrees of freedom. The model splits the
difference between the measurements of the total cross section at
$\sqrt s=1800$ GeV (see \fig{fig:1sigtot}). From \fig{fig:1rho}, we
note that the fit to $\rho$ is anchored at $\sqrt s=550$ GeV by the
very accurate measurement \cite{ua42} of UA4/2 and passes through the
E710 point \cite{E710rho}. The statistical uncertainty of the fitted
parameters is such that at 25 GeV the cross section predictions are
statistically uncertain to $\approx 1.3$\%, at 500 GeV are uncertain
to $\approx 1.6$\%, and at 2000 GeV are uncertain to $\approx 2.5$\%.

\begin{figure}[htbp] 
\begin{center}
\mbox{\epsfig{file=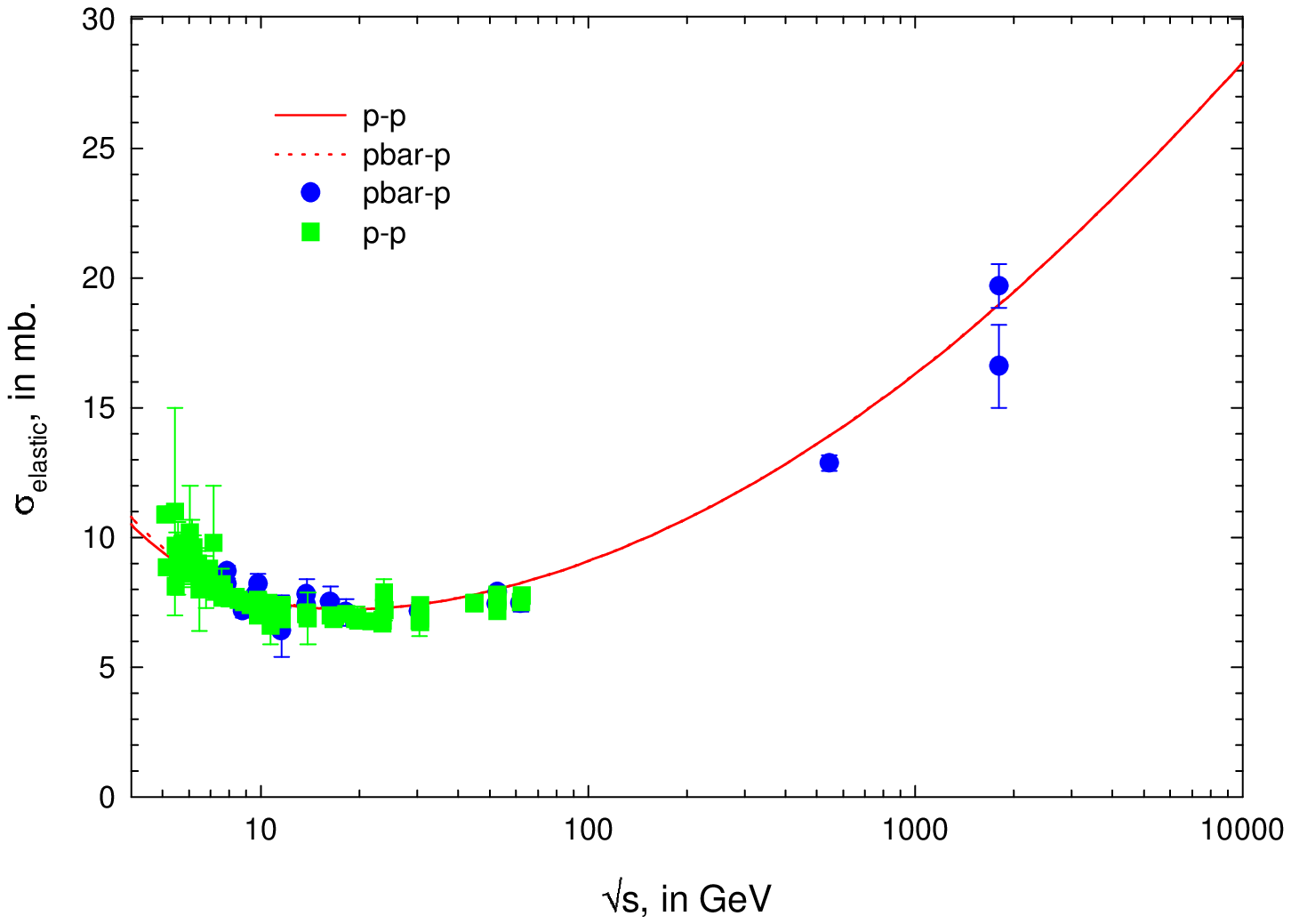,width=3.5in,%
bbllx=100pt,bblly=358pt,bburx=537pt,bbury=660pt,clip=}}
\end{center}
\caption[]{Elastic scattering cross sections for $pp$
and $p\bar p$ scattering. The solid line and squares are for $pp$ and
the dotted line and circles are for $\bar p p$.}
\label{fig:1sigel}
\end{figure}

In \fig{fig:1sigel} we show the prediction for the elastic cross
section along with the data for both $\bar pp$ and $pp$. The agreement
is excellent.  We note that $\sigma_{elastic}$ is rising more sharply
with energy than the total cross section $\sigma_{tot}$. Comparing
\fig{fig:1sigtot} with \fig{fig:1sigel}, we see that the ratio of the
elastic to total cross section is rising with energy.  The ratio is,
of course, bounded by the value for the black disk
\cite{bc,bcasymptopia}, {\em i.e.,} 0.5, as the energy goes to
infinity.

\begin{figure}[htbp] 
\begin{center}
\mbox{\epsfig{file=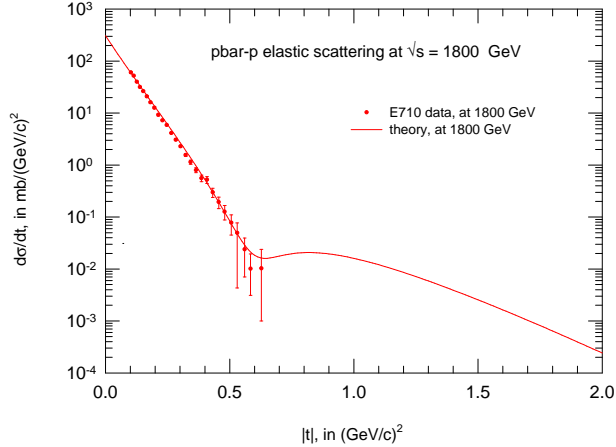,width=3.5in,%
bbllx=90pt,bblly=250pt,bburx=580pt,bbury=610pt,clip=}}
\end{center}
\caption[]{The elastic differential scattering cross
section for the reaction $\bar pp\rightarrow\bar pp$ at $\sqrt
s=1800$ GeV. The data points are from E710.}
\label{fig:1ds1800}
\end{figure}

Having fixed all parameters specifying our eikonal, we calculate
$d\sigma/dt$, for various values of $\sqrt s$. The differential cross
section at the Tevatron ($\sqrt{s}=1800$ GeV) is shown in
\fig{fig:1ds1800} 
along with E710 \cite{E710slope} data.  The
agreement over 4 decades is striking.  

\begin{figure}[htbp] 
\begin{center}
\mbox{\epsfig{file=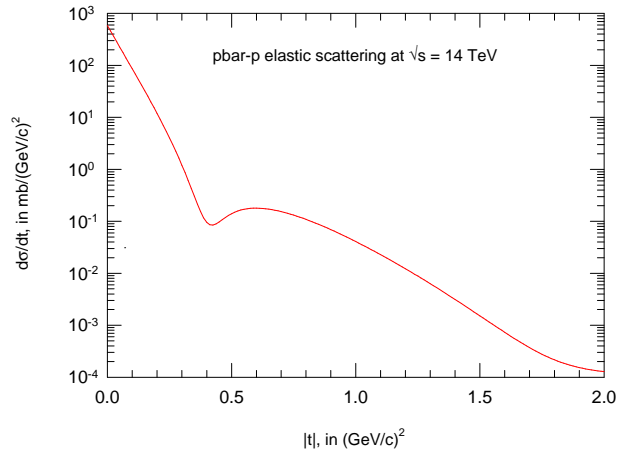,width=3.5in,%
bbllx=90pt,bblly=250pt,bburx=580pt,bbury=610pt,clip=}}
\end{center}
\caption[]{The elastic differential scattering
cross section for the reaction $\bar
pp\rightarrow\bar pp$ at LHC.}
\label{fig:1ds14000}
\end{figure}

Our prediction for the differential cross section
at $\sqrt s=14$ TeV, the energy of the LHC, is plotted in
\fig{fig:1ds14000}.
In particular, at small $|t|$, we predict that the
curvature parameter $C$ ($d\sigma/dt\propto e^{Bt+Ct^2}$ for small $t$; see ref. \cite{bc} for
details) is negative.  For energies much lower than 1800 GeV, the
observed curvature has been measured as positive. For 1800 GeV, we see
from \fig{fig:1ds1800} that the curvature parameter $C$ is compatible
with being zero.  Block and Cahn \cite{bc,bcasymptopia} have pointed
out that the curvature is predicted to go through zero near the
Tevatron energy and that it should become negative
thereafter. Asymptotically the proton approaches a black disk. Its
curvature is always negative \cite{bc,bcasymptopia}, $C=-R^4/192$,
where $R$ is the radius of the disk. Thus, the curvature has to pass
through zero as the energy increases.  `Asymptopia' is the energy
region (energies much larger than the Tevatron) where the
scattering approaches that of a sharp disk.

With the parameters we obtained from our fit, the total
cross section at the LHC (14 TeV) is predicted to be $\sigma_{tot}=108.0\pm 3.4$
mb, where the error is due to the statistical errors of the fitting
parameters.
\section{Photon-proton reactions}
\label{sec:gammap}

We assume that the photon behaves like a two quark system when it
interacts strongly. We therefore obtain $\gamma p$ scattering
amplitudes by performing the substitutions $\sigma_{ij} \rightarrow
\textstyle\frac{2}{3} \sigma_{ij}$ and $\mu_i \rightarrow
\sqrt{\textstyle\frac{3}{2}} \mu_i$ in the even eikonal for
nucleon--nucleon scattering, so that
\begin{equation}
\chi^{\gamma p}(s,b) =
i\left [\textstyle\frac{2}{3} \sigma_{qq}(s)
W\left(b;\sqrt{\textstyle\frac{3}{2}}\mu_{qq}\right)
+ \textstyle\frac{2}{3} \sigma_{qg}(s)
W\left(b;\sqrt{\textstyle\frac{3}{2}\mu_{qq}
\mu_{gg}}\right)
+ \textstyle\frac{2}{3} \sigma_{gg}(s)
W\left(b;\sqrt{\textstyle\frac{3}{2}}
\mu_{gg}\right)\right ]\, .
\label{chigammap}
\end{equation}

Using vector dominance, the photon-proton total cross section is then
written as
\begin{equation}
\sigma_{tot}^{\gamma p}(s)=
2P_{had}\int\,\left\{1-e^{-\chi_{I}^{\gamma p} (b,s)}
\cos[\chi_{R}^{\gamma p}(b,s)]\right\}\,d^2\vec{b}\, ,
\label{sigtotgammap}
\end{equation}
where $P_{had}$ is the probability that a photon interacts as a
hadron. We use the value $P_{had}=1/240$. This value is found by
normalizing the total $\gamma p$ cross section to the low energy data,
and is very close to that derived from vector dominance,
1/249. Using $f_{\rho}^2/4\pi=2.2$, $f_{\omega}^2/4\pi=23.6$ and
$f_{\phi}^2/4\pi=18.4$, we find $\Sigma_{V}(4\pi\alpha/f_V^2)=1/249$,
where $V=\rho,\omega,\phi$ (see Table XXXV, pag.\ 393 of Ref.\
\cite{bauer}).

\begin{figure}[htbp] 
\begin{center}
\mbox{\epsfig{file=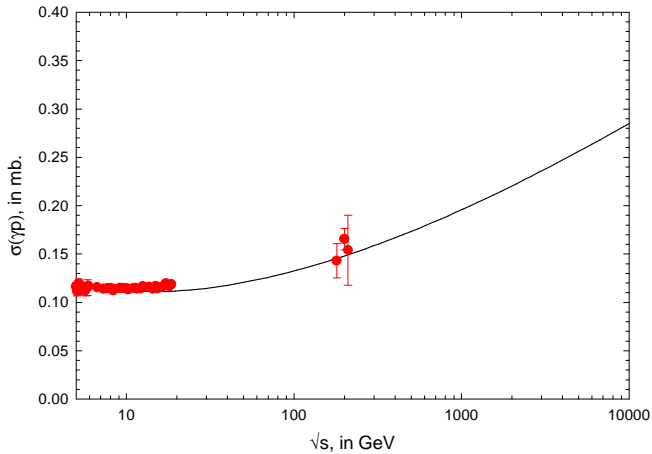,width=3.5in,%
bbllx=100pt,bblly=358pt,bburx=537pt,bbury=660pt,clip=}}
\end{center}
\caption[]{The total cross section for $\gamma p$
scattering.}
\label{fig:2sigtot}
\end{figure}

With all eikonal parameters fixed by the nucleon-nucleon data, we can
now calculate $\sigma_{tot}^{\gamma p}(s)$. The result is
shown in \fig{fig:2sigtot}. 
It reproduces the rising cross
section for $\gamma p$, using the parameters fixed by nucleon-nucleon
scattering. This prediction only uses the 9
parameters of the even eikonal, of which but 4 are important in the
upper energy region. The accuracy of our predictions are $\sim 1.5\%$,
from the statistical uncertainty in our eikonal parameters.

We next consider the `elastic' scatterings
\begin{eqnarray}
\gamma+p&\rightarrow &\rho_{virtual}+p\rightarrow\rho + p \; ,
\nonumber\\
\gamma+p&\rightarrow &\omega_{virtual}+p\rightarrow\omega + p \; ,
\nonumber\\
\gamma+p&\rightarrow &\phi_{virtual}+p\rightarrow\phi + p \; .
\label{Vpreactions}
\end{eqnarray}
Here the photon virtually transforms into a vector meson which
elastically scatters off of the proton. The strengths of these
reactions is ${\cal O}(\alpha)$ times a strong interaction cross
section.  The true elastic cross section is given by Compton
scattering on the proton, $\gamma+p\rightarrow\gamma+p$, which we can
visualize as
\begin{eqnarray}
\gamma+p&\rightarrow &\rho_{virtual}+p\rightarrow\rho
+ p \rightarrow \gamma+p \; ,
\nonumber\\
\gamma+p&\rightarrow &\omega_{virtual}+p\rightarrow\omega
+ p \rightarrow \gamma+p \; ,
\nonumber\\
\gamma+p&\rightarrow &\phi_{virtual}+p\rightarrow\phi
+ p \rightarrow \gamma+p \; .
\label{Compton}
\end{eqnarray}
It is clearly ${\cal O}(\alpha^2)$ times a strong interaction cross
section, and hence is much smaller than `elastic' scattering of
\eq{Vpreactions}.  Thus, we justify the use of \eq{sigtotgammap} to
calculate the total cross section, since only reactions with a photon
in the final state are neglected.

We evaluate $\rho$ and the slope $B$ for the `elastic' scattering
expressed in \eq{Vpreactions}, with $\rho$ and $B$ being the same for all 3 reactions.

\begin{figure}[htbp] 
\begin{center}
\mbox{\epsfig{file=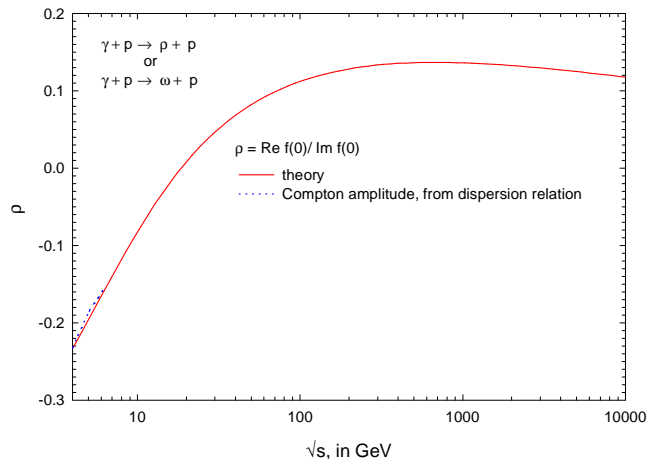,width=3.5in,%
bbllx=100pt,bblly=358pt,bburx=537pt,bbury=660pt,clip=}}
\end{center}
\caption[]{Ratio of the
real to imaginary part of the forward scattering amplitude for the
`elastic' reactions $\gamma +p\rightarrow V_i + p$, where $V_i$ is
$\rho^0$, $\omega^0$ or $\phi^0$.  The dotted curve is for Compton
scattering  from dispersion relations \protect\cite{gilman}.
It has been slightly displaced from the solid curve for clarity in viewing.}
\label{fig:2rhovec}
\end{figure}

The dependence of $\rho$ with the energy is shown in
\fig{fig:2rhovec}.  Damashek and Gilman \cite{gilman} have calculated
the $\rho$ value for Compton scattering on the proton using
dispersion relations, {\em i.e.,} the true elastic scattering reaction
for photon-proton scattering.  We compare this calculation, the dotted
line in \fig{fig:2rhovec},
with our prediction of $\rho$ (the solid
line).  The agreement is so close that we had to move the two curves
apart so that they may be viewed more clearly.

\begin{figure}[htbp]
\begin{center}
\mbox{\epsfig{file=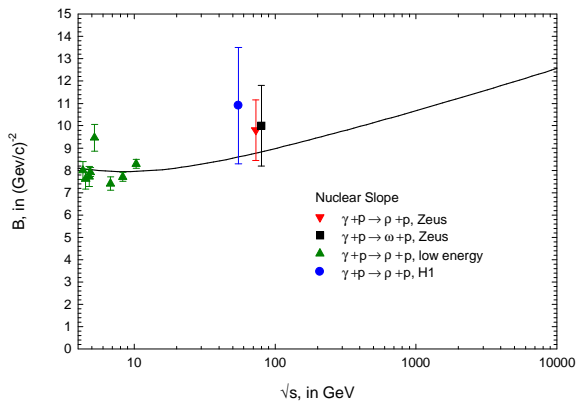,width=3.5in,%
bbllx=67pt,bblly=394pt,bburx=500pt,bbury=685pt,clip=}}
\end{center}
\caption[]{Nuclear slope parameter
for the `elastic' reaction $\gamma +p\rightarrow V_i + p$, where $V_i$
is $\rho^0$, $\omega^0$ or $\phi$. For the reaction $\gamma
+ p\rightarrow\rho^0 + p$, the inverted triangles are the Zeus data,
the circles are the H1 data, and the triangles are the low energy data.
For the reaction $\gamma + p\rightarrow\omega^0 + p$, the squares are
the Zeus data.}
\label{fig:2bvec}
\end{figure}

In \fig{fig:2bvec} 
we show our results for the slope $B$
as a function of the energy.  The available experimental data for
`elastic' $\rho p$ and $\omega p$ final states are also
plotted. Again, the agreement of theory and experiment is very good.

To calculate the elastic cross sections $\sigma_{elastic}^{Vp}$ and
differential cross sections $d\sigma^{Vp}/dt$ as a function of
energy, we use
\begin{equation}
\sigma_{elastic}^{Vp}(s) = P_{had}^{Vp}
\int\left|1-e^{i\chi^{\gamma p}(b,s)}\right|^2\,d^2\vec{b},
\label{sigelgp}
\end{equation}
where $P_{had}^{Vp}$ is the appropriate probability for a photon
to turn into $V$, with $V=\rho,\ \omega$ or $\phi$. The differential
scattering cross section is given by
\begin{equation}
\frac{d\sigma^{Vp}}{dt}(s,t)=\frac{P_{had}^{Vp}}{4\pi}\left|
\int J_0(qb)(1-e^{i\chi^{\gamma p}(b,s)})\,d^2\vec{b}\,\right|^2,
\label{dsdt2gp}
\end{equation}
where $t=-q^2$.

Since we normalize our data to the cross section found with
$\chi^{\gamma p}$, and not to
$(\sigma^{\pi^+}_{tot}+\sigma^{\pi^-}_{tot})/2$,
we must multiply all $f_{V}^2/4\pi$ by 1.65.
Hence, our effective couplings are
$f_{\rho\,\rm eff}^{2}/4\pi=3.6$,
$f_{\omega\,\rm eff}^2/4\pi=38.9$, and
$f_{\phi\,\rm eff}^2/4\pi=30.4$.

\begin{figure}[htbp] 
\begin{center}
\mbox{\hspace{0.2in}\epsfig{file=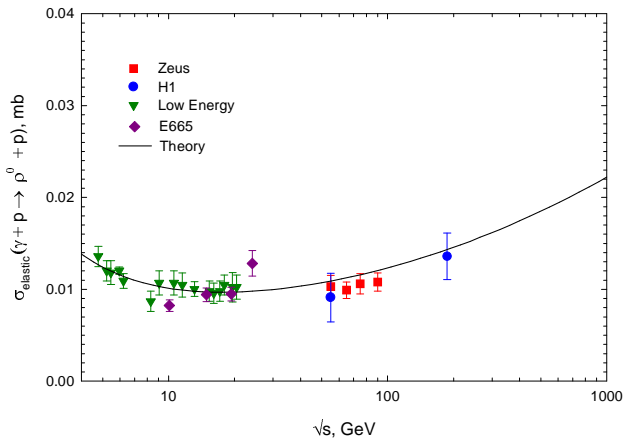,width=3.5in,%
bbllx=80pt,bblly=330pt,bburx=540pt,bbury=650pt,clip=}}
\end{center}
\caption[]{The `elastic' photoproduction cross section,
for the reaction $\gamma +p\rightarrow\rho^0 + p$. The squares are
Zeus data, the circles are H1 data, and the inverted triangles the low
energy data.}
\label{fig:2sigrho}
\end{figure}

\begin{figure}[htbp] 
\begin{center}
\mbox{\hspace{0.2in}\epsfig{file=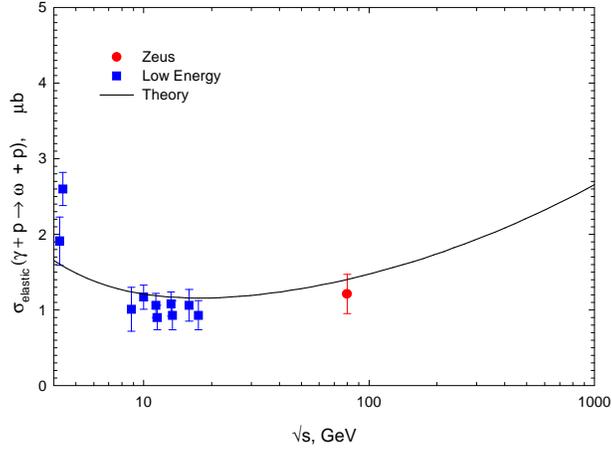,width=3.5in,%
bbllx=90pt,bblly=250pt,bburx=537pt,bbury=560pt,clip=}}
\end{center}
\caption[]{The `elastic' photoproduction cross section
for the reaction $\gamma +p\rightarrow\omega^0 + p$. The circles are
Zeus data, and the squares are the low energy data. }
\label{fig:2sigomeg}
\end{figure}

Our evaluation of the `elastic' cross section for the reactions
$\gamma +p\rightarrow\rho^0 + p$ and $\gamma +p\rightarrow\omega^0 +
p$ are shown in Figs.\ \ref{fig:2sigrho} and \ref{fig:2sigomeg},
respectively. 

\begin{figure}[htp] 
\begin{center}
\mbox{\epsfig{file=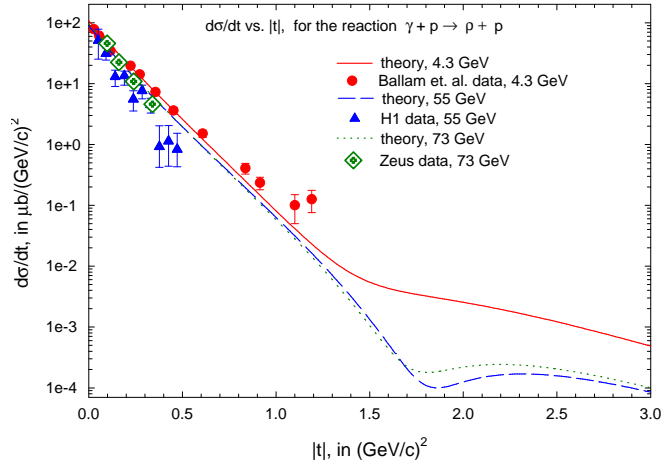,width=3.5in,%
bbllx=100pt,bblly=240pt,bburx=530pt,bbury=550pt,clip=}}
\end{center}
\caption[]{The differential cross
section for the `elastic' reaction $\gamma +p\rightarrow \rho^0+ p$.
The solid curve and the circles (Ballam {\em et al.} data)
are at $\sqrt s$= 4.3 GeV, the dashed curve and triangles (H1 data)
are at $\sqrt s$= 55 GeV, and the dotted curve and diamonds are at
$\sqrt s$= 73 GeV (Zeus data).}
\label{fig:2dsrho}
\end{figure}

\begin{figure}[htp]
\begin{center}
\mbox{\epsfig{file=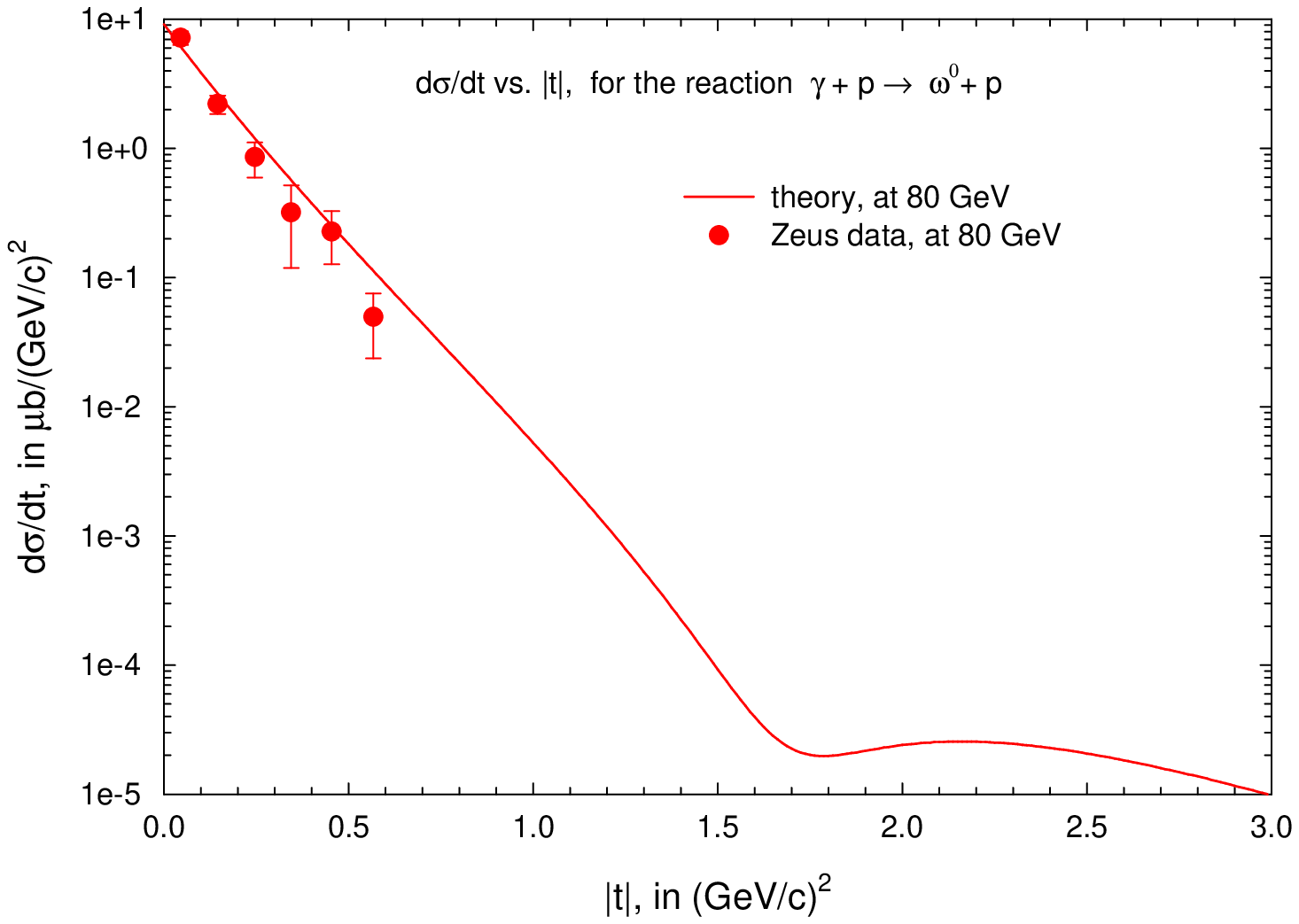,width=3.5in,%
bbllx=80pt,bblly=240pt,bburx=510pt,bbury=550pt,clip=}}
\end{center}
\caption[]{The differential cross
section for the `elastic' reaction $\gamma +p\rightarrow
\omega^0+ p$ at $\sqrt s$=80 GeV.  The circles are the Zeus data. }
\label{fig:2dsomega}
\end{figure}

\begin{figure}[htbp]
\begin{center}
\mbox{\epsfig{file=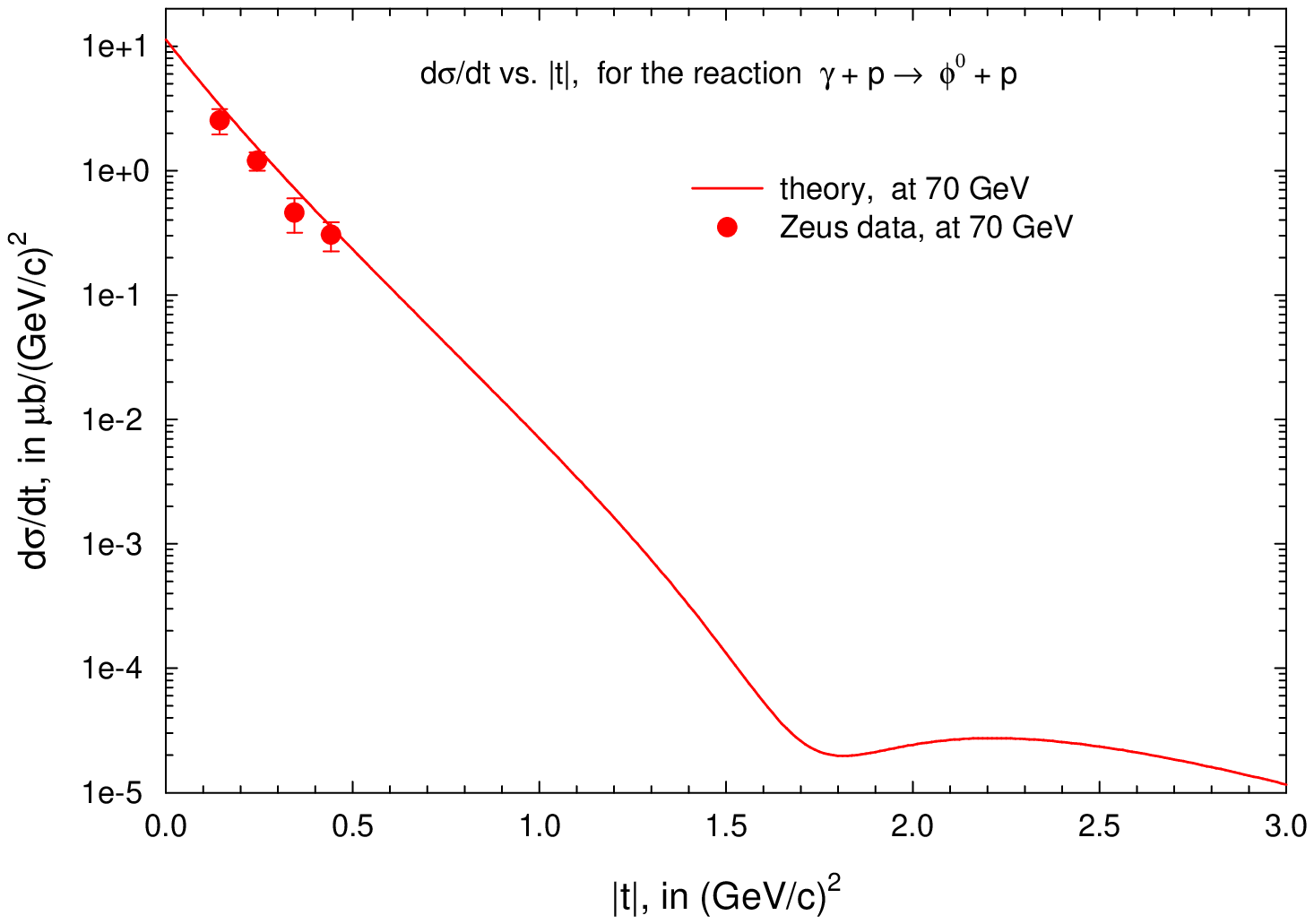,width=3.5in,%
bbllx=40pt,bblly=225pt,bburx=520pt,bbury=570pt,clip=}}
\end{center}
\caption[]{The differential cross
section for the `elastic' reaction $\gamma +p\rightarrow
\phi^0+ p$ at $\sqrt s$=70 GeV.  The circles are the Zeus data. }
\label{fig:2ds70phi}
\end{figure}

The differential cross section, $d\sigma/dt$, for the `elastic'
reactions $\gamma +p\rightarrow \rho^0+ p$, $\gamma +p\rightarrow
\omega^0+ p$ and $\gamma +p\rightarrow \phi^0+ p$ are plotted in
Figs.\ \ref{fig:2dsrho}, \ref{fig:2dsomega}, and
\ref{fig:2ds70phi}, respectively. 
The agreement, in absolute
normalization and shape, of our results for all three light vector
mesons with the experimental data for all available energies
reinforces our confidence in the model.
\section{Photon-Photon Interactions}
\label{gammagamma}

\begin{figure}[htbp]
\begin{center}
\mbox{\epsfig{file=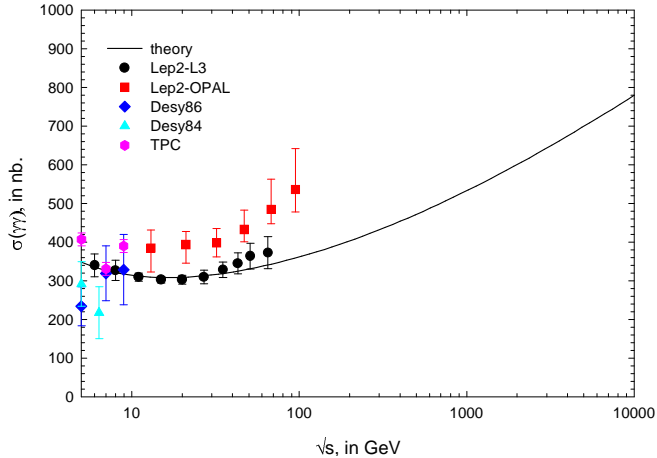,width=3.5in,%
bbllx=100pt,bblly=358pt,bburx=537pt,bbury=660pt,clip=}}
\end{center}
\caption[]{The total cross section for $\gamma
\gamma$ scattering. The data sources are indicated in the legend.}
\label{fig:3sigtot}
\end{figure}

In this section, we consider $\gamma\gamma$ interactions. As it was
done for $\gamma p$ interactions, we will start from the eikonal
$\chi^{\gamma p}(s,b)$ and multiply every cross section by 2/3
and multiply each $\mu$ by $\sqrt {3/2}$. Therefore,
\begin{equation}
\chi^{\gamma \gamma}(s,b) =
\textstyle i\left[ \frac{4}{9} \sigma_{qq}(s)
W\left(b;\frac{3}{2}\mu_{qq}\right)
+ \frac{4}{9} \sigma_{qg}(s)W\left(b;\frac{3}{2}
\sqrt{\mu_{qq}\mu_{gg}} \right)
+ \frac{4}{9} \sigma_{gg}(s)W\left(b;\frac{3}{2}\mu_{gg} \right)\right
]\, .
\label{chigammagamma}
\end{equation}
Using vector dominance we obtain,
\begin{equation}
\sigma_{tot}^{\gamma \gamma}(s) =
2P_{had}^2 \int\,
\left\{1-e^{-\chi_{I}^{\gamma \gamma}(b,s)}
\cos[\chi_{R}^{\gamma p}(b,s)] \right\} \, d^2\vec{b},
\label{sigtotgammagamma}
\end{equation}
where $P_{had}=1/240$ is the probability that a photon will
interact as a hadron.  In \fig{fig:3sigtot} 
we plot our results for
$\sigma_{tot}^{\gamma \gamma}(s)$ as a function of the
energy, and compare it to the various sets of
experimental data. We note that our prediction fits the L3 data, but doesn't fit the OPAL results.
\section{Proton-air cross sections}
 Cosmic ray experiments measure the penetration in the atmosphere
 of particles with energies in excess of those accelerated by existing
 machines---interestingly, their energy range covers the energy of the
 Large Hadron Collider (LHC) and extends beyond it. However, extracting
 proton--proton cross sections from cosmic ray observations is far from
 straightforward~\cite{gaisser}. By a variety of experimental techniques,
 cosmic ray experiments map the atmospheric depth at which cosmic ray
 initiated showers develop.
 The measured shower attenuation length ($\Lambda_m$) is not only
 sensitive to the interaction length of the protons in the atmosphere
 ($\lpa$), with
\begin{equation}
\Lambda_m = k \lpa = k { 13.5 m_p \over \spai} \,,  \label{eq:Lambda_m}
\end{equation}
 but also depends on the rate at which the energy of the primary proton
 is dissipated into electromagnetic shower energy observed in the
 experiment. The latter effect is parameterized in Eq.\,(\ref{eq:Lambda_m})
 by the parameter $k$; $m_p$ is the proton mass and $\spai$ the inelastic
 proton-air cross section. The value of $k$ depends on the inclusive
 particle production cross section in nucleon and meson interactions
 on the light nuclear target of the atmosphere and its energy dependence.
 We here ignored the fact that particles in the cosmic  ray "beam" may be
 nuclei, not just protons. Experiments allow for this by omitting
 from their analysis showers which dissipate their energy high in the
 atmosphere, a signature that the initial energy is distributed over the
 constituents of a nucleus.

\begin{figure}[htbp]
\begin{center}
\mbox{\epsfig{file=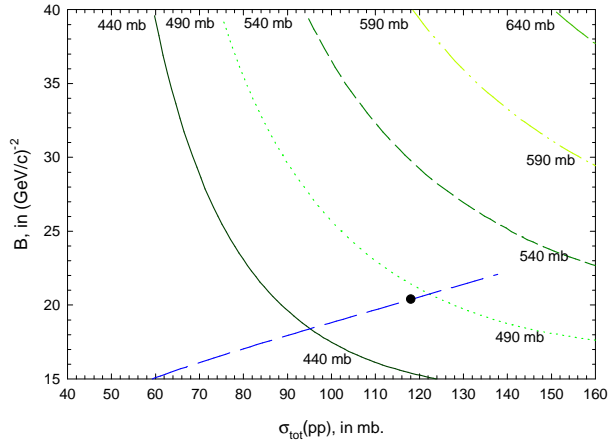%
              ,width=3.5in,bbllx=100pt,bblly=358pt,bburx=537pt,bbury=660pt,clip=%
}}
\end{center}
\caption[]{  $B$ dependence on the pp total cross section $\sigma_{pp}$. The
 five curves are lines  of constant  $\spai$,  of 440, 490, 540, 590 and
 640 mb---the central value is the published Fly's Eye value, and the others
 are $\pm 1\sigma$ and $\pm 2\sigma$. The dashed curve is a plot of our
 QCD-inspired fit of $B$ against $\sigma_{pp}$.  The dot is our value for
 $\sqrt s=30$ TeV, the Fly's Eye energy.}
\label{fig:p-air}
\end{figure}

 The extraction of the pp cross section from the cosmic ray data is a two
 step process. First, one calculates the $p$-air total cross section from
 the measured inelastic cross section
\begin{equation}
\spai = \spa - \spae - \spaqe \,.  \label{eq:spa}
\end{equation}
 Next, the Glauber method\cite{yodh} is used to transform the measured
 value of $\spai$ into a proton--proton total cross section $\sigma_{pp}$;
 all the necessary steps are calculable in the theory. In Eq.\,(\ref{eq:spa})
 the measured cross section for particle production is supplemented with
 the elastic and quasi-elastic cross section, as calculated by the
 Glauber theory, to obtain the total cross section $\spa$. The subsequent
 relation between $\spai$ and $\sigma_{pp}$ involves the slope of the
 forward scattering amplitude for elastic $pp$ scattering,
\begin{equation}
B = \left[ {d\over dt} \left(\ln{d\sigma_{pp}^{\rm el}\over dt}\right)
 \right]_{t=0} \,,
\end{equation}
%
 and is shown in Fig.\,\ref{fig:p-air}, 
which plots $B$ against
 $\sigma_{pp}$, for 5 curves of different values of $\spai$.
 This summarizes the reduction procedure
 from $\spai$ to $\sigma_{pp}$~\cite{gaisser}.
 Also plotted in Fig.\,\ref{fig:p-air} is a curve of $B$ {\em vs.}
 $\sigma_{pp}$ which will be discussed later.
	
 A significant drawback of the method is that one needs a model of
 proton--air interactions to complete the loop between the measured
 attenuation length $\Lambda_m$ and the cross section $\spai$,
 {\em i.e.,} the value of $k$ in Eq. (\ref{eq:Lambda_m}). We minimize the impact of theory 
by using our QCD-inspired parameterization of the forward
 proton--proton and proton--antiproton scattering amplitudes
 which is analytic, unitary and {\em simultaneously} fits all data of $\sigma_{\rm tot}$, $B$
 and $\rho$. Using vector meson
 dominance and the additive quark models, we have shown that it  accommodates a wealth of
 data on photon-proton and photon-photon interactions without the
 introduction of new parameters. Because the model is
 both unitary and analytic, it has high energy predictions that are
 essentially theory--independent.  In particular, it also
 {\em simultaneously} fits $\sigma_{pp}$ and $B$, forcing a relationship
 between the two. Specifically, the $B$ {\em vs.} $\sigma_{pp}$ prediction
 of the model is shown as the dashed curve in Fig.\,\ref{fig:p-air}. 
The
 dot corresponds to our prediction of $\sigma_{pp}$ and $B$ at
 $\sqrt s$ = 30 TeV. It is seen to be slightly below the curve for
 490 mb, the lower limit of the Fly's Eye measurement, which was
 made at $\sqrt s\approx$ 30 TeV.

\begin{figure}[htbp]
\begin{center}
\mbox{\epsfig{file=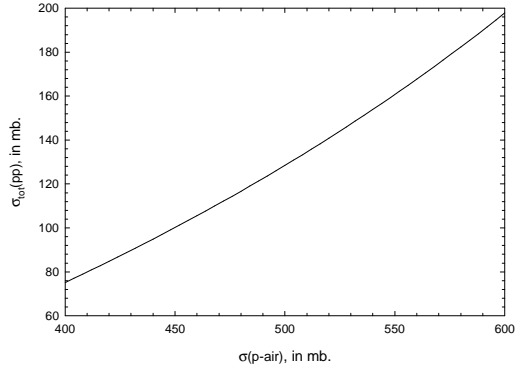%
              ,width=3.5in,bbllx=55pt,bblly=345pt,bburx=570pt,bbury=680pt,clip=%
}}
\end{center}
\caption[]{A plot of the predicted total pp cross section $\sigma_{pp}$, in mb
 {\em vs.} the measured p-air cross section, $\spai$, in mb.
}
\label{fig:sigpp_p-air}
\end{figure}

 In Fig.\,\ref{fig:sigpp_p-air}, 
we have plotted the values of
 $\sigma_{pp}$ {\em vs.} $\spai$ that are deduced from the
 intersections of the $B$-$\sigma_{pp}$ curve  with the $\spai$
 curves of Fig.\,\ref{fig:p-air}. Figure~\,\ref{fig:sigpp_p-air}
 allows the conversion of the measured $\spai$ to $\sigma_{pp}$ .

\begin{figure}[htbp]
\begin{center}
\mbox{\epsfig{file=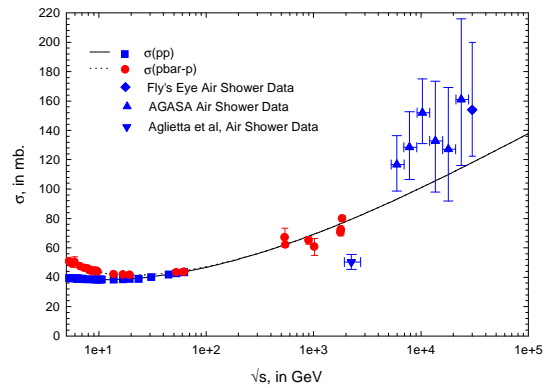%
              ,width=3.5in,bbllx=85pt,bblly=340pt,bburx=570pt,bbury=680pt,clip=%
}}
\end{center}
\caption[]{\protect
{  A plot of the QCD-inspired fit of the total nucleon-nucleon cross section
 $\sigma_{pp}$, in mb {\em vs.} $\sqrt s$, in Gev. The cosmic ray data that
 are shown have been converted from $\spai$ to $\sigma_{pp}$ using the
 results of Fig.~\ref{fig:sigpp_p-air}.
}
}
\label{fig:sigtodorpp}
\end{figure}

 Our prediction for the total cross section $\sigma_{pp}$ as a
 function of energy is confronted with all of the accelerator and
 cosmic ray measurements\cite{fly,akeno,eastop} in Fig.\,\ref{fig:sigtodorpp}.
 For inclusion in  Fig.\,\ref{fig:sigtodorpp}, we have  calculated
 the cosmic ray values of $\sigma_{pp}$ from the
 {\em published} experimental values of $\spai$, using the results
 of Fig.\,\ref{fig:sigpp_p-air}. We note the predicted curve is systematically lower than the cosmic ray points, roughly about the level of one standard deviation.

 It is at this point important to recall Eq.\,(\ref{eq:Lambda_m})
 and consider the fact that the extraction of  $\spai$ from the
 measurement of $\Lambda_m$ requires a determination of the parameter
 $k$. The measured depth $X_{\rm max}$ at which a shower reaches
 maximum development in the atmosphere, which is the basis of the
 cross section measurement in Ref.~\cite{fly}, is a combined measure
 of the depth of the first interaction, which is determined by
 the inelastic cross section, and of the subsequent shower development,
 which has to be corrected for. The position of $X_{\rm max}$ also
 directly affects the rate of shower attenuation with atmospheric depth
 which is the alternative procedure for extracting $\spai$.

 The model dependent rate of shower development and its fluctuations
 are the origin of the deviation of $k$ from unity
 in Eq.\,(\ref{eq:Lambda_m}). Its values range from 1.5 for a model
 where the inclusive cross section exhibits Feynman scaling, to 1.1
 for models with large scaling violations\cite{gaisser}. The comparison
 between data and experiment in Fig.\,\ref{fig:sigtodorpp} is further
 confused by the fact  that the AGASA\cite{akeno} and Fly's Eye\cite{fly}
 experiments used different values of $k$ in the analysis of their data,
 {\em i.e.,} AGASA used $k=1.5$ and Fly's Eye used $k=1.6$.

\begin{figure}[h]
\begin{center}
\mbox{\epsfig{file=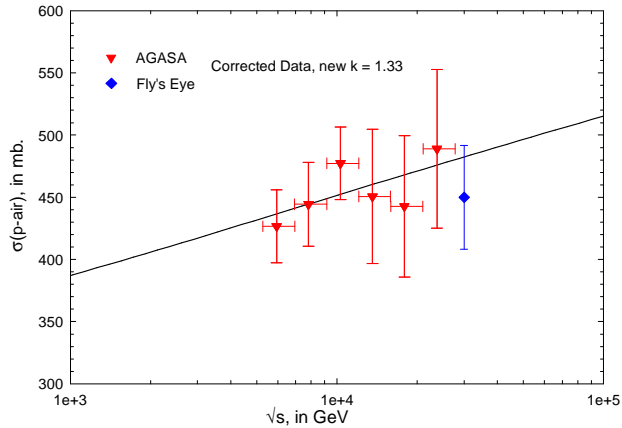%
		  ,width=3.5in,bbllx=86pt,bblly=375pt,bburx=514pt,bbury=664pt,clip=%
}}
\end{center}
\caption[]{\protect
{  A $\chi^2$ fit of the measured AGASA and Fly's Eye data for $\spai$, in mb,
 as a function of the energy, $\sqrt s$, in GeV. The result of the fit for the
 parameter $k$ in Eq. (\ref{eq:Lambda_m}) is $k=1.33\pm0.04$.
}
}
\label{fig:p-aircorrected3}
\end{figure}

 We therefore decided to match the data to our prediction and extracted
 a common value for $k=1.33 \pm 0.04$. This neglects the possibility
 that $k$ may show a weak energy dependence over the range measured.
 In Fig.\,\ref{fig:p-aircorrected3}
we have replotted the high energy
 cosmic ray data for our prediction of $\spai$ {\em vs.} $\sqrt s$,
 with the common value of $1.33$ obtained from a $\chi^2$ fit. Clearly,
 we have an excellent fit, with good agreement between AGASA and Fly's Eye.
 The analysis gives  $\chi^2=1.75$ for 6 degrees of freedom (the low
 $\chi^2$ is probably due to overestimates of experimental errors).
 This result for $k$ is interesting---it is close to the value of $1.2$
 obtained using the SIBYLL simulation\cite{sybill} for inclusive particle
 production. This represents a consistency check in the sense that our
 model for forward scattering amplitudes and SIBYLL share the same
 underlying physics. The increase of the total cross section with
 energy to a black disk of soft partons is the shadow of increased
 particle production which is modeled by the production of (mini)-jets
 in QCD. The difference between the $k$ values of 1.20 and 1.33
 could be understood because the experimental measurement integrates
 showers in a relatively wide energy range, which tends to increase
 the value of $k$.

In the near term, we look  forward
 to the possibility of repeating this analysis with the higher
 statistics of the HiRes~\cite{HiRes} cosmic ray experiment that
 is currently in progress and the Auger~\cite{Auger} Observatory.

In conclusion, we have successfully united the high energy cross section results ($\sqrt s\approx 30$ TeV) of the cosmic ray measurements with the accelerator cross section measurements, under a common rubric, the QCD-inspired analysis.
%
%
%

\end{document}